\documentclass[twocolumn,trackchanges]{aastex701}

\usepackage{amsmath}
\usepackage{orcidlink}
\usepackage[version=4]{mhchem}
\usepackage[utf8]{inputenc}
\usepackage{newunicodechar}
\newunicodechar{∼}{\textasciitilde}

\newcommand{\revisedtext}[1]{\textcolor{black}{#1}}

\begin{document}

\title{The Role of Formation Location in Shaping Sulfur-, Nitrogen-, and Carbon-Bearing Species in Super-Earth and Sub-Neptune Atmospheres}

\shortauthors{Werlen et al.}
\correspondingauthor{Aaron Werlen}

\author[orcid=0009-0005-1133-7586, sname='Werlen']{Aaron Werlen}
\affiliation{Institute for Particle Physics and Astrophysics, ETH Zurich, CH-8093 Zurich, Switzerland}
\affiliation{Department of Earth, Planetary, and Space Sciences, University of California, Los Angeles, CA 90095, USA}
\email[show]{werlen@ucla.edu}

\author[orcid=0000-0002-9020-7309, sname="Burn"]{Remo Burn}
\affiliation{Observatoire de la Côte d’Azur, Université Cote d’Azur, Laboratoire Lagrange, 06300 Nice, France}

\email{remo.burn@oca.eu}

\author[orcid=0000-0001-6110-4610, sname='Dorn']{Caroline Dorn}
\affiliation{Institute for Particle Physics and Astrophysics, ETH Zurich, CH-8093 Zurich, Switzerland}
\email{dornc@ethz.ch}

\author[orcid=0009-0004-1292-3969, sname='Felix']{Lukas Felix}
\affiliation{Department of Space Research and Technology, Technical University of Denmark, DK-2800 Kgs. Lyngby, Denmark}
\email{lukfe@dtu.dk}

\author[orcid=0009-0007-2917-2390, sname='Salmi']{Annika Salmi}
\affiliation{Institute for Particle Physics and Astrophysics, ETH Zurich, CH-8093 Zurich, Switzerland}
\email{asalmi@student.ethz.ch}

\begin{abstract}
Atmospheric compositions of sub-Neptunes and super-Earths are often interpreted as tracers of formation location relative to volatile ice lines. However, prolonged magma oceans can chemically equilibrate with primordial atmospheres and modify accreted volatile signatures. In this study, we couple a synthetic planet population from the Bern Generation III formation model to an extended global chemical equilibrium framework including sulfur and nitrogen chemistry, and compare accreted and equilibrated compositions for $\sim 1200$ \revisedtext{young} planets \revisedtext{shortly after formation ($\sim 4$ Myr)} formed inside and outside the water ice line. We find that interior–atmosphere equilibration systematically alters elemental ratios and molecular abundances. The atmospheric C/O ratio shifts relative to the accreted state and remains systematically higher for planets formed outside the ice line. Nitrogen-bearing species (\ce{NH3}, \ce{N2}) are strongly depleted through dissolution into the silicate melt, while minor amounts of \ce{HCN} are produced, leading to low atmospheric nitrogen abundances. Sulfur-bearing species remain more abundant than nitrogen-bearing species; during equilibration, accreted \ce{H2S} partitions into the interior and small amounts of \ce{SO2} form, but overall sulfur abundances depend only weakly on formation location. Silicon-bearing gases (\ce{SiH4}, \ce{SiO}) are generated in substantial amounts, with narrower distributions for planets formed outside the ice line. We identify atmospheric C/O, \ce{SiH4}, and \ce{H2O} as potential indicators of formation location, while nitrogen depletion emerges as a generic outcome of magma ocean equilibration. Comparison with characterized sub-Neptunes such as TOI-270 d, K2-18 b, and GJ 3470 b shows broad consistency with oxygen-dominated, metal-rich atmospheres shaped by interior–atmosphere exchange.
\end{abstract}



\section{Introduction}

The James Webb Space Telescope (JWST) has opened a new window into the chemical diversity of sub-Neptune atmospheres, with detections of \ce{H2O}, \ce{CH4}, \ce{CO}, \ce{CO2}, and other species now becoming increasingly common \citep[e.g.,][]{madhusudhan_carbon-bearing_2023,benneke_jwst_2024,schmidt_comprehensive_2025,felix_competing_2025,Davenport_toi421b_2025}. This growing observational access to atmospheric chemistry has renewed interest in using atmospheric composition to constrain planet formation. In particular, the carbon-to-oxygen (C/O) ratio has long been interpreted as a tracer of formation location relative to volatile ice lines in protoplanetary disks \citep{oberg_effects_2011}. \revisedtext{More generally, accreted material is often assumed to preserve the compositional imprint of its formation history, with the final atmospheric composition reflecting the relative contributions of gas, volatile-rich solids, and refractory material \citep[e.g.,][]{venturini_nature_2020}.} While variations in phase partitioning have been explored \citep[e.g.,][]{burn_radius_2024}, chemical reprocessing through reactions between the interior and atmosphere has received comparatively little attention. For giant planets, this framework further implies that observed atmospheric abundances primarily reflect the composition of accreted nebular gas and late solid enrichment, whereas material accreted prior to runaway gas accretion remains largely hidden \citep{shibata_giant_planet_metallicity_2022}.

However, super-Earths and sub-Neptunes differ fundamentally from gas giants in their interior structure and thermal evolution. Even a few weight percent of hydrogen and helium are sufficient to maintain long-lived magma oceans at the atmosphere--magma ocean interface (AMOI) \citep[e.g.,][]{lopez_understanding_2014, ginzburg_super-earth_2016, misener_importance_2022}. Under these conditions, chemical equilibrium between the molten silicate interior, the metal phase, and the overlying atmosphere can substantially redistribute volatile elements \citep[e.g.,][]{kite_superabundance_2019, kite_atmosphere_2020, lichtenberg_redox_2021, dorn_hidden_2021, schlichting_chemical_2022, young_differentiation_2025, misener_atmospheres_2023, seo_role_2024, luo_interior_2024, cherubim_oxidation_2025, lichtenberg_constraining_2025, werlen_atmospheric_2025, werlen_sub-neptunes_2025, steinmeyer_coupled_2026}. As a result, key atmospheric diagnostics such as the C/O ratio and the \ce{H2O} inventory need not reflect primordial disk abundances, but instead emerge from global interior--atmosphere equilibration \citep{werlen_atmospheric_2025, werlen_sub-neptunes_2025}.

Recent time-dependent models coupling interior evolution and atmospheric chemistry have further demonstrated that magma ocean equilibration can substantially modify atmospheric C/O ratios and water inventories during planetary evolution, but these studies have largely focused on carbon- and oxygen-bearing species \citep{steinmeyer_coupled_2026}. This naturally raises a broader question: if major tracers such as C/O and water can be strongly modified by deep equilibration, to what extent can atmospheric compositions still preserve an imprint of formation location? In particular, nitrogen-, sulfur-, and carbon-bearing species may encode additional information about bulk composition and volatile accretion histories \citep{pacetti_giantcompo_2022,crossfield_sulfur_2023}. Whether such signatures survive partitioning into silicate and metal phases, or are erased by global chemical redistribution during equilibration, remains largely unexplored.

In this study, we couple a planetary population synthesis model to the global chemical equilibrium framework of \citet{grimm_new_2026}, extended to include carbon partitioning \citep{werlen_atmospheric_2025} as well as nitrogen and sulfur species in multiple phases. \revisedtext{We evaluate the chemical equilibrium of a synthetic population of sub-Neptunes at a single time step shortly after disk dispersal ($\sim 4$ Myr), without modeling their subsequent evolution, and compare the accreted and equilibrated compositions for planets formed inside and outside the water ice line.} By analyzing the resulting atmospheric abundances and elemental ratios, we assess whether systematic differences between formation regions persist after magma ocean–atmosphere equilibration. Our results quantify the degree to which atmospheric compositions are reshaped by interior processing and identify molecular patterns that may serve as tracers of the formation environment in the JWST era.

This paper is structured as follows. In Section~\ref{sec:methods}, we describe the chemical network and the adopted synthetic planet population. In Section~\ref{sec:results}, we quantify the effects of interior–atmosphere equilibration on elemental ratios and key atmospheric species, including the C/O ratio, nitrogen-, sulfur-, carbon-, and silicon-bearing species, as well as the atmospheric mass fraction, atmospheric metal mass fraction, and mean molecular weight. In Section~\ref{sec:discussion}, we place our results in the context of previous work, compare them to available observations, and discuss implications and limitations. Section~\ref{sec:conclusion} summarizes our main findings.

\section{Methods}
\label{sec:methods}

\subsection{Chemical Thermodynamics}
\label{sec:chemical_network}

We model interior–atmosphere equilibration using the Global Chemical Equilibrium (GCE) code by \citet{grimm_new_2026}, which based on the formulation of \citet{schlichting_chemical_2022}. The model computes equilibrium states across three coexisting phases—metal, silicate melt, and gas—by simultaneously enforcing chemical equilibrium, mass balance, and elemental conservation.

Relative to \citet{schlichting_chemical_2022}, the chemical network employed here is substantially expanded. Carbon partitioning into the metal phase is included following \citet{werlen_atmospheric_2025}. In addition, sulfur chemistry is treated self-consistently in all phases, with \ce{H2S} and \ce{SO2} in the gas phase, \ce{FeS} and \ce{FeSO4} in the silicate melt, and elemental sulfur in the metal phase. We further include nitrogen-bearing species, allowing for \ce{N2}, \ce{HCN}, and \ce{NH3} in the gas phase, as well as dissolved \ce{N2} in the silicate melt. All nitrogen- and sulfur-bearing species and their associated partitioning represent new additions introduced in this work.

Equilibrium abundances of all phase components, together with the total number of moles in each phase, are obtained using the numerical scheme detailed \citet{grimm_new_2026}. The full reaction network, equilibrium formulation, and governing equations are provided in Appendix~\ref{ap:chem_network}. Additionally, Appendix~\ref{ap:therm_data} documents the thermodynamic data sources and discusses the assumptions regarding ideality.

\revisedtext{Helium is not included as a phase component in the chemical equilibrium framework because we do not model helium solubility or partitioning between the gas, silicate, and metal reservoirs. Although the initial helium abundance is known from the initially accreted H/He envelope, the final atmospheric helium abundance cannot be determined self-consistently within the present model. All reported gas-phase abundances and derived atmospheric quantities therefore refer to the chemically active gas phase only.}

As in previous studies \citep{schlichting_chemical_2022, young_phase_2024, werlen_atmospheric_2025, werlen_sub-neptunes_2025, werlen_effects_2026}, we define the astrophysical core as consisting of both metal and silicate phases. This definition reflects the possibility that the partitioning of light elements into the metal phase can reduce its density to values comparable to those of the silicate phase. As a result, the formation of a distinct core–mantle boundary is inhibited. Experimental and ab-initio studies support this behavior under high-pressure conditions \citep{hirao_compression_2004, terasaki_hydrogen_2009, tagawa_experimental_2021, li_earths_2020, luo_interior_2024, young_differentiation_2025}.

\subsubsection{Hydrogen Solubility Treatment}
\label{sec:h2_solubility}

\revisedtext{The partitioning of \ce{H2} between the gas and silicate phases depends sensitively on the adopted hydrogen solubility law, with strong consequences for the atmospheric mass fraction and the redox state of the atmosphere \citep{grimm_new_2026}. Here, we fix the equilibrium constant for the \ce{H2} solubility reaction,
\begin{equation}
    k_{\rm eq} = \frac{x_{\ce{H2},s}}{f_{\ce{H2},g}},
\end{equation}
where $x_{\ce{H2},s}$ is the mole fraction of dissolved \ce{H2} in the silicate melt and $f_{\ce{H2},g}$ is the fugacity of atmospheric \ce{H2}. In this formulation, the abundance of dissolved \ce{H2} increases with the fugacity, and therefore pressure, of atmospheric \ce{H2}. This contrasts with the fixed partition-coefficient approach,
\begin{equation}
    k_{\rm D} = \frac{x_{\ce{H2},s}}{x_{\ce{H2},g}},
\end{equation}
used in \citet{schlichting_chemical_2022, werlen_atmospheric_2025, werlen_sub-neptunes_2025}, in which the ratio between dissolved and gaseous \ce{H2} is fixed. The two prescriptions should be viewed as limiting cases rather than uniquely established descriptions of hydrogen solubility. The fixed-$k_{\rm eq}$ formulation adopted here promotes more efficient hydrogen storage in the silicate melt at high pressure and therefore tends to reduce the atmospheric mass fraction relative to the fixed-$k_{\rm D}$ formulation. Recent experimental and theoretical studies support the possibility of substantial hydrogen incorporation into silicate melts under sub-Neptune conditions \citep{young_phase_2024, young_differentiation_2025, miozzi_experiments_2025, horn_building_2025, gilmore_coreenvelope_2026, young_influences_2026}, but the pressure and temperature dependence of this process remains uncertain. We therefore regard the hydrogen-solubility prescription as one of the main model assumptions in this work, and we will explore its impact further in future studies.}
\subsection{Initial Planet Population}
\label{sec:population}

\begin{figure}
    \centering
    \includegraphics{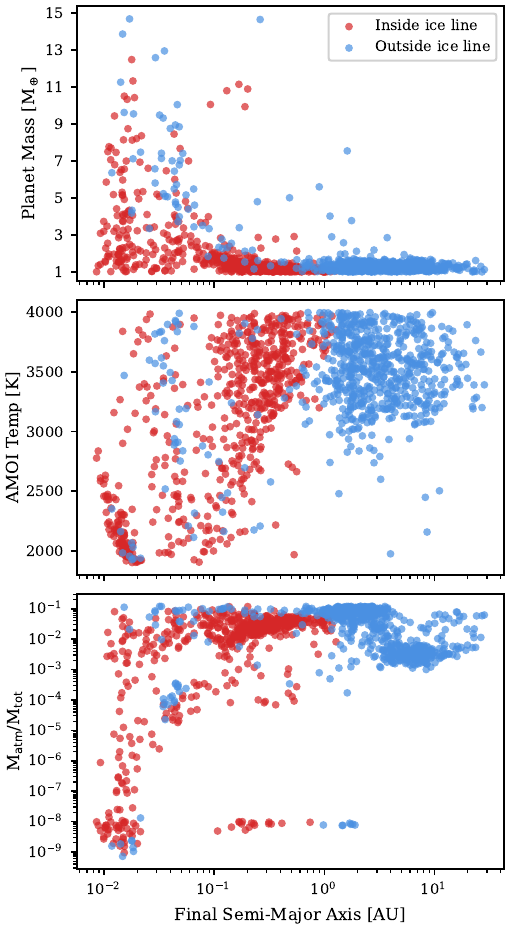}
    \caption{Final semi-major axis after disk-driven migration as a function of planetary mass and atmosphere--magma ocean interface (AMOI) temperature for the initial population. The population is separated into planets formed inside and outside the water ice line, with the inside--ice-line sample defined by an accreted water mass fraction $\leq 5\,\mathrm{wt}\%$. The two populations span comparable planetary mass ranges, indicating no strong mass bias between the inside-- and outside--ice-line samples. In contrast, the AMOI temperature distributions differ systematically with formation location, with temperatures below $\sim 3000\,\mathrm{K}$ occurring predominantly for planets formed inside the ice line. \revisedtext{The bottom panel shows the initial atmospheric mass fraction, $M_{\rm atm}/M_{\rm tot}$. Low atmospheric mass fractions occur almost exclusively among planets formed inside the ice line and are correlated with the lower AMOI temperatures in this population.}}    
    \label{fig:sma_mass}
\end{figure}

\begin{figure*}
    \centering
    \includegraphics[width=1\textwidth]{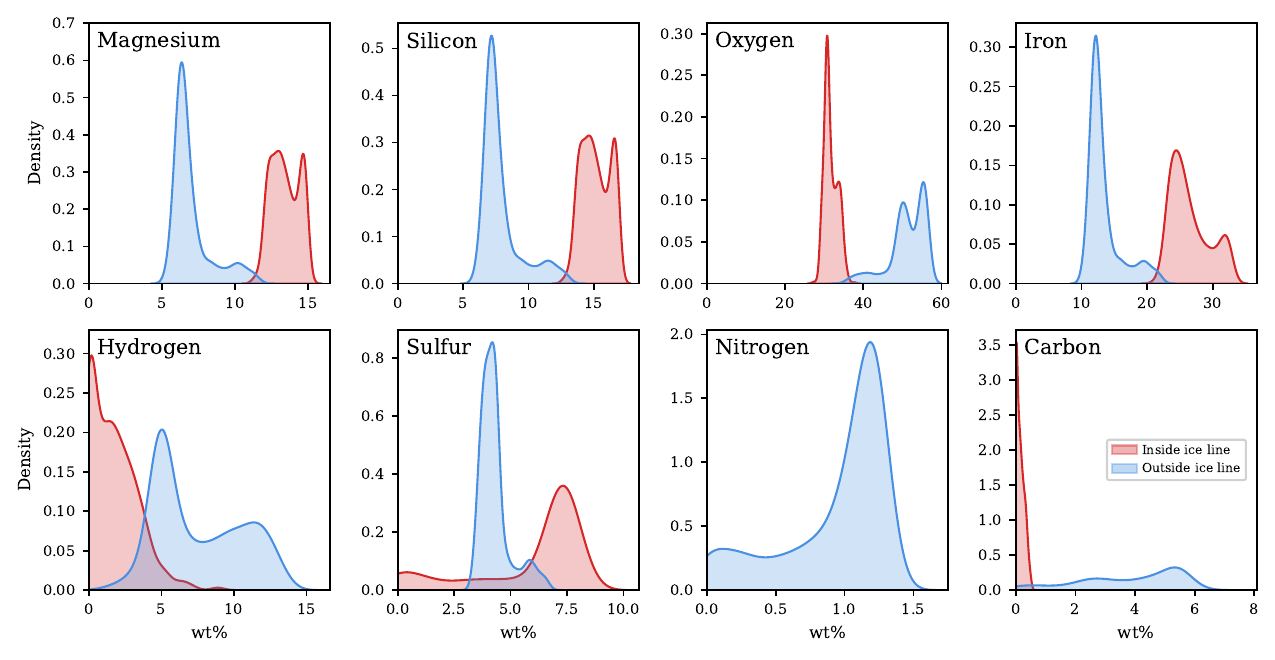}
    \caption{Kernel density estimates of the bulk elemental mass fractions of the initial planet population. The population is separated into planets that formed inside and outside the water ice line. Planets formed outside the ice line exhibit systematically higher oxygen and hydrogen mass fractions, reflecting enhanced accretion of volatile-rich material, which in turn leads to a relative depletion of refractory elements such as magnesium, iron, sulfur, and silicon. High bulk nitrogen and carbon abundances are exclusively associated with planets that formed outside the ice line.}
    \label{fig:kde_bulk}
\end{figure*}

We adopt a synthetic population of young sub-Neptunes from the New Generation Planetary Population Synthesis (NGPPS), generated using the Generation~III Bern global model of planet formation and evolution. The model and its implementation are identical to those used in \citet{werlen_sub-neptunes_2025}, and we therefore provide only a brief overview here. Detailed descriptions of the model architecture and numerical methods are given by \citet{emsenhuber_new_2021} and \citet{emsenhuber_planetary_2021}. The population used here was calculated for Solar-mass stars with initially 50 embryos placed randomly following a log-uniform distribution in the disk. Recent updates of the planets' evolution were discussed by \citet{burn_water-rich_2024, burn_radius_2024}, where the detailed extraction of the state of the planet after disk dissipation used here is discussed in \citet{burn_water-rich_2024}.

\revisedtext{Within the NGPPS framework, the elemental inventories of each planet are inherited from the solids and gas accreted along its formation and migration history. The disk composition is constructed separately for refractory and volatile material. This prescription represents a reset-composition scenario, in which the composition of planetesimals is determined by the local disk conditions at their formation locations, rather than being inherited unchanged from the presolar cloud \citep[e.g.,][]{pacetti_giantcompo_2022}. Refractory minerals are assigned following the equilibrium condensation calculations of \citet{thiabaud_stellar_2014}, adopting Solar elemental abundances from \citet{lodders_solar_2003}. Volatile carriers are prescribed using the organic-free model of \citet{marboeuf_stellar_2014,marboeuf_planetesimals_2014}, where the relative abundances of major ice species are constrained by interstellar and cometary measurements. The volatile inventory includes \ce{H2O}, \ce{CO}, \ce{CO2}, \ce{CH4}, \ce{CH3OH}, \ce{NH3}, \ce{N2}, and \ce{H2S}, which provide the initial sources of H, O, C, N, and S in the population. At each orbital distance, the local disk pressure and temperature determine which species are condensed and can be incorporated into planetesimals. These solid compositions are calculated at the initialization of the viscously and irradiation-heated disks, using initial conditions informed by Class I protoplanetary disks \citep{tychoniec_vla_2018, emsenhuber_planetary_2023}. Planetesimals are assumed to remain at their formation locations; radial drift, migration, and scattering of planetesimals are not included. The cumulative accretion of solid material, including both refractory and volatile components, together with nebular hydrogen gas, then sets the bulk elemental abundances used as input to the equilibrium calculations.}

The planetary population is extracted at the time of disk dispersal ($\sim$~4~Myr, varying with disk properties), when planets are still young and hot and are expected to host global magma oceans. For each planet, we extract the total planetary mass, bulk elemental composition, and the atmosphere–magma ocean interface (AMOI) temperature. The AMOI temperature is obtained from the one-dimensional hydrostatic atmospheric structure calculations within the NGPPS model, where it corresponds to the temperature at the base of the accreted atmosphere at disk dispersal. These quantities define the initial conditions for the chemical equilibrium calculations described in Section~\ref{sec:chemical_network}.

We construct our planet sample by selecting objects with total masses between $1$ and $15\,\mathrm{M_\oplus}$ and atmosphere--magma ocean interface (AMOI) temperatures between $1900$ and $4000\,\mathrm{K}$, corresponding to the temperature range assumed for molten planetary interiors. In addition, we exclude planets with atmospheric mass fractions exceeding $12\,\mathrm{wt}\%$, where the atmosphere is defined as the combined hydrogen and helium mass. This cutoff is chosen in accordance with \citet{rogers_unveiling_2021}, who found that the occurrence rate of sub-Neptunes drops significantly above atmospheric mass fractions of $\sim 10\,\mathrm{wt}\%$.

Relative to \citet{werlen_sub-neptunes_2025}, which was restricted to planets with masses above $2\,\mathrm{M_\oplus}$, the present sample additionally includes lower-mass planets, thereby increasing the overall population size. In contrast, the earlier study did not impose a cutoff in atmospheric mass fraction, which resulted in a small number of planets with very massive atmospheres that are not included here. 

The final initial sample considered in this work comprises 1260 planets, compared to only a few hundred in the previous study. Of these, 1177 converge to stable equilibrium solutions.

Convergence is evaluated using the squared sum of the residuals of the coupled non-linear system of equations described in Appendix~\ref{ap:chem_network} and \citet{grimm_new_2026}. A solution is considered converged if this residual metric falls below $10^{-3}$, although including solutions with larger residuals does not alter the qualitative population-level trends presented below.

In contrast to \citet{werlen_sub-neptunes_2025}, which focused on hydrogen-, oxygen-, and carbon-bearing species, the present study additionally extracts sulfur and nitrogen inventories from the population synthesis output and includes \ce{SiH4} in the equilibrium calculation. These inventories serve as inputs for the extended chemical network employed here and enable a self-consistent treatment of sulfur and nitrogen partitioning during interior–atmosphere equilibration. The distribution of initial elemental abundances across the population is discussed in detail below.

Throughout this paper, we distinguish between planets that predominantly formed inside and outside the water ice line. Planets with an accreted water mass fraction $\leq 5\,\mathrm{wt}\%$ are classified as forming inside the water ice line, while planets exceeding this threshold are classified as forming exterior to the water ice line.

\subsubsection{Properties of the Initial Planet Population}

Figure~\ref{fig:sma_mass} provides an overview of the orbital, mass, and thermal properties of the initial population by showing the final semi-major axis after disk-driven migration as a function of planetary mass and AMOI temperature, with color indicating the formation location. Planets in both formation groups span comparable mass ranges across the sampled orbital distances, indicating that the inside-- and outside--ice-line populations are not systematically biased in planetary mass. In contrast, the AMOI temperatures differ markedly between the two groups, with lower interface temperatures (below $\sim 3000\,\mathrm{K}$) occurring predominantly for planets formed inside the ice line. This difference is primarily driven by variations in hydrogen accretion, which is higher for planets formed outside the ice line. \revisedtext{The initial atmosphere mass fractions, here defined as hydrogen plus helium, shown in the lower panel of Figure~\ref{fig:sma_mass} directly illustrate this trend, with low-atmosphere-mass planets occurring almost exclusively in the inside--ice-line population.} This results from the temperature dependence of dust opacities, which regulates atmospheric cooling and therefore the rate of gas accretion \citep{lee_cool_2015, coleman_situ_2017}. Figure~\ref{fig:kde_bulk} summarizes the bulk elemental compositions of the population using kernel density estimates of elemental mass fractions. Compositional differences are evident between the two formation groups: planets formed exterior to the ice line show enhanced abundances of oxygen, hydrogen, carbon, and nitrogen, while refractory elements such as magnesium, silicon, iron, and sulfur are correspondingly depleted. These trends reflect the accretion of volatile-rich ices beyond the ice line.
 
\section{Results}
\label{sec:results}

\begin{figure}
    \centering
    \includegraphics{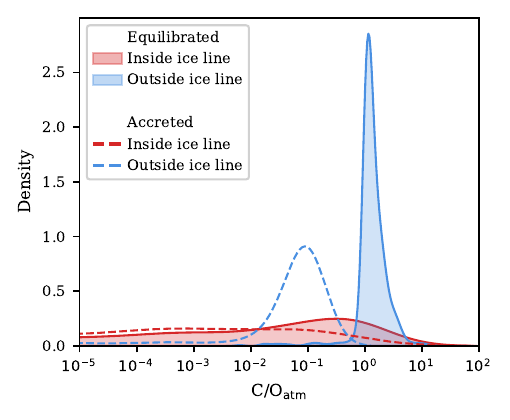}
    \caption{Kernel density estimates of atmospheric C/O ratios for the accreted composition (dashed curves) and after interior--atmosphere equilibration (filled distributions), separated into planets formed inside and outside the water ice line. Planets formed outside the ice line occupy a comparatively narrow C/O range both before and after equilibration, but the equilibrated distribution is shifted to higher C/O. In contrast, planets formed inside the ice line exhibit a broader (less tightly constrained) C/O distribution in both the accreted and equilibrated cases.}    
    \label{fig:kde_c_o}
\end{figure}

\begin{figure*}
    \centering
    \includegraphics[width=1\textwidth]{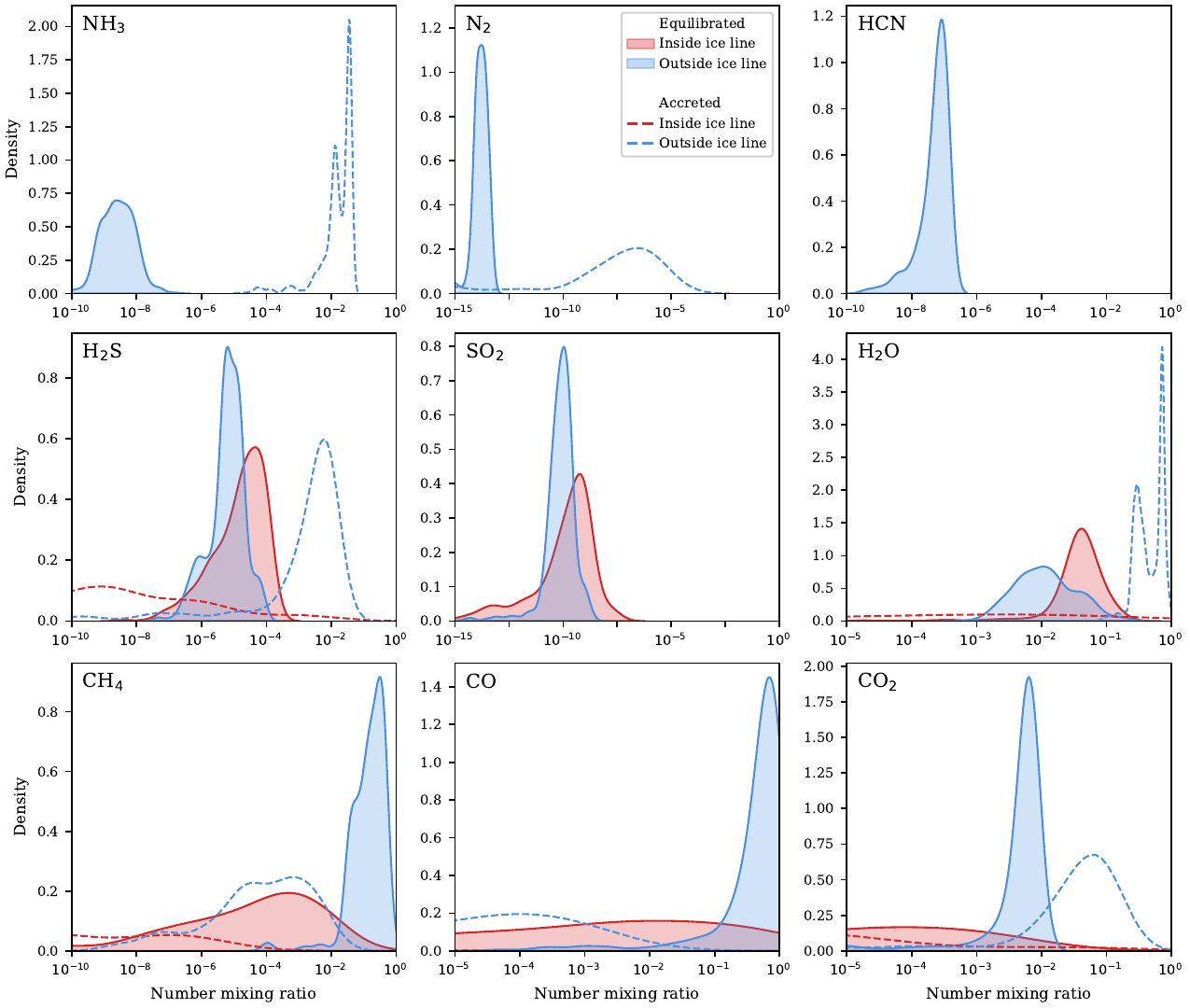}
    \caption{Kernel density estimates of atmospheric number mixing ratios for selected species, comparing the accreted compositions (dashed curves, where present) with the equilibrated interior--atmosphere compositions (filled distributions). The population is separated into planets formed inside and outside the water ice line. Interior--atmosphere equilibration substantially reshapes the atmospheric composition. Nitrogen-bearing species are strongly depleted relative to the accreted state: \ce{NH3} and \ce{N2} decrease in abundance, while small amounts of \ce{HCN} are produced. For planets formed outside the ice line, equilibration enhances \ce{CH4} and \ce{CO} compared to their accreted values, whereas \ce{CO2} becomes less abundant. In contrast, planets formed inside the ice line exhibit broader and more uniformly distributed carbon-bearing abundances after equilibration. \ce{H2S} and \ce{SO2} display similar equilibrated distributions for both formation environments, while \ce{H2O} retains clear differences between planets formed inside and outside the ice line.}   
    \label{fig:kde_atmosphere}
    \end{figure*}

\begin{figure}
    \centering
    \includegraphics{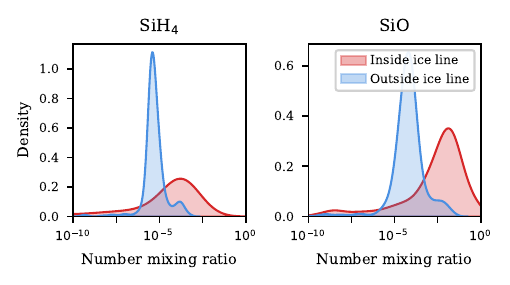}
    \caption{Kernel density estimates of atmospheric \ce{SiH4} and \ce{SiO} number mixing ratios after interior--atmosphere equilibration, separated into planets formed inside and outside the water ice line. Planets formed outside the ice line occupy a comparatively narrow range of \ce{SiH4} and \ce{SiO} abundances, whereas planets formed inside the ice line exhibit a broader and less tightly constrained \ce{SiH4} distribution.}   
    \label{fig:kde_sih4}
\end{figure}

\begin{figure*}
    \centering
    \includegraphics[width=1\textwidth]{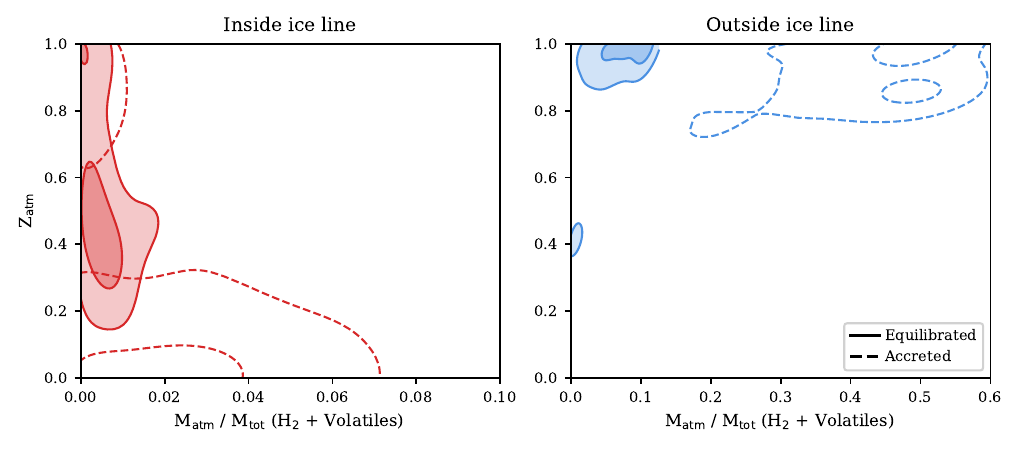}
    \caption{Two-dimensional kernel density estimates of the atmospheric metal mass fraction (Z$_\text{atm}$), defined as the total mass fraction of all atmospheric species except hydrogen, as a function of the total atmosphere mass fraction (M$_\text{atm}$/M$_\text{tot}$) (hydrogen plus volatile species). Dashed contours show the accreted composition, while filled contours indicate the equilibrated interior--atmosphere state. Only the $1\sigma$ and $2\sigma$ contour levels are shown. The population is separated into planets formed inside and outside the water ice line. Planets formed outside the ice line exhibit similar Z$_\text{atm}$ distributions before and after equilibration, indicating only modest compositional modification. In contrast, planets formed inside the ice line develop an additional high-Z$_\text{atm}$ peak after equilibration, reflecting enhanced metal enrichment through interior--atmosphere exchange. In both formation groups, the total atmospheric mass fraction decreases after equilibration due to volatile partitioning into the interior.}
    \label{fig:Zatm}
\end{figure*}

\begin{figure}
    \centering
    \includegraphics{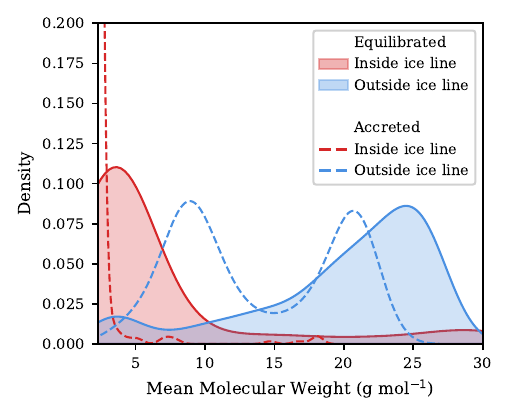}
    \caption{Kernel density estimates of the atmospheric mean molecular weight. Dashed contours show the accreted composition, computed under the assumption that all accreted volatiles reside in the gaseous atmosphere. Filled contours indicate the equilibrated interior--atmosphere state. Interior--atmosphere exchange shifts the mean molecular weight distribution toward higher values.}    \label{fig:kde_mean_molecular_weight}
\end{figure}

\subsection{Atmospheric C/O Ratio}

Figure~\ref{fig:kde_c_o} shows the atmospheric C/O ratio for planets formed inside and outside the water ice line, comparing the accreted and equilibrated compositions.

Planets formed inside the ice line exhibit broad C/O distributions both before and after equilibration. In contrast, planets formed outside the ice line attain systematically higher C/O ratios after equilibration.

These trends are directly linked to the molecular abundance patterns shown in Figure~\ref{fig:kde_atmosphere}. The carbon- and oxygen-bearing species \ce{H2O}, \ce{CH4}, \ce{CO}, and \ce{CO2} display distinct distributions between the two formation groups after equilibration. Planets formed inside the ice line are enriched in \ce{H2O} and depleted in carbon-bearing species relative to planets formed outside the ice line. This behavior reflects the strong depletion of bulk carbon in the inside--ice-line population (see Figure~\ref{fig:kde_bulk}), which limits the atmospheric abundances of \ce{CH4}, \ce{CO}, and \ce{CO2} even after equilibration.

Conversely, planets formed outside the ice line accrete substantially larger bulk carbon inventories, together with enhanced hydrogen fractions. These conditions favor the formation of carbon species, particularly \ce{CH4}. The combined increase in carbon-bearing species and comparatively lower \ce{H2O} abundances shifts the equilibrated atmospheres toward higher C/O ratios.

\subsection{Atmospheric Sulfur and Nitrogen Species}

Figure~\ref{fig:kde_atmosphere} shows the distribution of nitrogen-, sulfur-, and oxygen-bearing atmospheric species for both formation locations.

Planets formed outside the ice line can accrete atmospheres with high initial \ce{NH3} mixing ratios. Interior--atmosphere equilibration subsequently depletes the atmosphere in nitrogen-bearing species such as \ce{NH3} and \ce{N2}, while producing \ce{HCN}. The overall low atmospheric nitrogen abundances after equilibration indicate that the vast majority of nitrogen partitions into the interior, where it is stored in the silicate melt.

A qualitatively similar but weaker effect is observed for sulfur-bearing species. During equilibration, \ce{H2S} decreases and small amounts of \ce{SO2} are produced. As for nitrogen, sulfur can partition into the silicate melt and the metal phase, which limits the atmospheric sulfur abundances after equilibration. The resulting \ce{H2S} and \ce{SO2} distributions are similar for planets formed inside and outside the ice line. This is caused by the opposing bulk trends of sulfur and hydrogen across the two formation environments (see Figure~\ref{fig:kde_bulk}): planets formed inside the ice line have higher bulk sulfur abundances but lower hydrogen abundances compared to the planets formed outside the ice line. These compensating trends lead to comparatively small differences in atmospheric \ce{H2S} between both populations. A similar compensation between sulfur and oxygen yields small differences in the \ce{SO2} distributions.

\subsection{Atmospheric Silicate Species}

Figure~\ref{fig:kde_sih4} shows the atmospheric mixing ratios of \ce{SiH4} and \ce{SiO} after equilibration. Planets formed outside the ice line occupy a relatively narrow range of \ce{SiH4} and \ce{SiO} mixing ratios at systematically lower values, whereas planets formed inside the ice line exhibit broader and more weakly constrained distributions.

Consequently, \ce{SiH4} does not follow the same trend as the carbon-bearing species \ce{CH4}, \ce{CO}, and \ce{CO2}, which are enhanced outside the ice line.

This contrasting behavior is again primarily driven by bulk composition. Planets formed outside the ice line are relatively depleted in refractory elements such as silicon (see Figure~\ref{fig:kde_bulk}), as their higher volatile content dilutes the refractory fraction. The reduced relative bulk silicon inventory hence limits the production of \ce{SiH4} and \ce{SiO}.

\subsection{Atmospheric Mass Fraction, Metal Mass Fraction, and Mean Molecular Weight}

Figure~\ref{fig:Zatm} shows the atmospheric metal mass fraction, $Z_{\mathrm{atm}}$, defined as the mass fraction of all atmospheric species excluding \ce{H2}, as a function of the atmosphere mass fraction, M$_{\mathrm{atm}}$/M$_{\mathrm{tot}}$. The accreted values are computed under the assumption that all volatile ices delivered during formation reside in the atmosphere as gaseous species. This assumption leads to the large atmospheric mass fractions inferred for the accreted population beyond the ice line. \revisedtext{As discussed in Section~\ref{sec:chemical_network}, helium is not included as a phase component; the atmospheric mass fractions, atmospheric metal mass fractions, and mean molecular weights reported here therefore refer to the chemically active gas phase only.}

For planets formed outside the ice line, the accreted and equilibrated $Z_{\mathrm{atm}}$ distributions largely overlap, while the atmosphere mass fractions are significantly reduced after equilibration due to the partitioning of volatiles into the interior.

In contrast, planets formed inside the ice line exhibit a pronounced redistribution in both $Z_{\mathrm{atm}}$ and atmosphere mass fraction. Objects that initially accrete atmospheres with very low $Z_{\mathrm{atm}}$ shift toward substantially higher values after equilibration, developing a peak at $Z_{\mathrm{atm}} \sim 0.4$--0.5. This increase reflects interior--atmosphere exchange, which enhances the abundance of metal-bearing species in the atmosphere. At the same time, their total atmosphere mass fractions are reduced, commonly reaching values of order $1\%$ or below.

These compositional changes are reflected in the atmospheric mean molecular weight (Figure~\ref{fig:kde_mean_molecular_weight}). For planets formed inside the ice line, the accreted atmospheres exhibit very low mean molecular weights, consistent with hydrogen-dominated compositions. After equilibration, the distribution shifts toward higher values, in line with the enhanced atmospheric metal mass fractions discussed above.

For planets formed outside the ice line, the accreted mean molecular weight distribution is bimodal. This bimodality primarily reflects differences in \ce{H2} accretion, with \ce{H2}-rich planets forming a low mean molecular weight population and \ce{H2}-poor planets retaining higher mean molecular weights. After equilibration, the distribution shifts toward higher mean molecular weights, consistent with the increased contribution of heavier species such as \ce{CO} and \ce{CO2}.

Overall, interior--atmosphere equilibration systematically reduces the total atmosphere mass fraction while increasing both the atmospheric metal mass fraction and the atmospheric mean molecular weight.

\section{Discussion}
\label{sec:discussion}

The population-level trends identified here challenge the common interpretation of atmospheric compositions as direct tracers of disk chemistry and formation location. If magma ocean–atmosphere equilibration systematically redistributes volatile elements and modifies elemental ratios, then observed atmospheric abundances encode both primordial accretion histories and subsequent interior processing. Interpreting atmospheric measurements therefore requires disentangling these coupled effects.

\subsection{C/O ratio as a formation location tracer}\label{sec:formation_location}

With JWST, retrieval studies of sub-Neptunes increasingly allow us to characterize carbon- and oxygen-bearing species to constrain atmospheric C/O ratios \citep[e.g.,][]{benneke_jwst_2024,schmidt_comprehensive_2025,felix_competing_2025}. Theoretical predictions of C/O ratios are hence becoming increasingly important.

In a previous study, \citet{werlen_atmospheric_2025} showed that atmospheric C/O ratios can vary by several orders of magnitude solely as a function of the planetary hydrogen mass fraction, even when the bulk refractory composition is held constant. \revisedtext{That study adopted a fixed-$k_{\rm D}$ hydrogen-solubility prescription, whereas the present work uses the fixed-$k_{\rm eq}$ treatment described in Section~\ref{sec:h2_solubility}. This difference affects how much hydrogen remains in the atmosphere after equilibration and therefore influences the atmospheric redox state and the resulting C/O ratio. In particular, the fixed-$k_{\rm eq}$ prescription allows more efficient dissolution of \ce{H2} into the silicate melt at high pressure, reducing the direct sensitivity of atmospheric C/O to the accreted hydrogen mass fraction compared to the fixed-$k_{\rm D}$ case. Despite this difference, the broader conclusion remains unchanged:} atmospheric composition is not directly inherited from the protoplanetary disk but instead emerges from interior--atmosphere equilibration. Similar trends were reported by \citet{seo_role_2024}. Overall, this picture remains valid. In the present study, we find that planets formed outside the ice line tend to exhibit systematically higher and more tightly constrained C/O ratios (see Figure~\ref{fig:kde_c_o}). Nevertheless, these equilibrated C/O ratios still differ from the accreted values.

\revisedtext{Our results are also consistent with recent thermal--chemical evolution models by \citet{steinmeyer_coupled_2026}, who adopted the same fixed-$k_{\rm eq}$ hydrogen-solubility treatment and found that atmospheric C/O is more sensitive to the accreted water abundance than to the accreted \ce{H2} inventory. This behavior is consistent with the efficient dissolution of \ce{H2} into the silicate melt under the fixed-$k_{\rm eq}$ prescription, which limits the effect of additional accreted hydrogen on atmospheric redox state.}

\subsection{Global Water Mass Fraction}\label{sec:water_mass_fraction}

In a previous study based on a smaller subset of the same synthetic population, \citet{werlen_sub-neptunes_2025} showed that planets formed inside the ice line exhibit enhanced atmospheric water mass fractions after equilibration, despite accreting less bulk water than planets formed outside the ice line.

Using the expanded chemical network presented here---including sulfur-, nitrogen-, and silicon-bearing species---and adopting \revisedtext{the fixed-$k_{\rm eq}$ hydrogen-solubility prescription discussed in Section~\ref{sec:h2_solubility}}, we recover the same qualitative trend \revisedtext{reported by \citet{werlen_sub-neptunes_2025}, despite the different treatment of \ce{H2} solubility}. Atmospheric number mixing ratios of \ce{H2O} remain systematically higher for planets formed inside the ice line compared to those formed outside.

\revisedtext{Compared to the fixed-$k_{\rm D}$ prescription used in \citet{werlen_sub-neptunes_2025}, the fixed-$k_{\rm eq}$ treatment adopted here allows more efficient dissolution of \ce{H2} into the silicate melt at high pressure.} This enhanced hydrogen storage promotes endogenic water formation in the interior and strengthens volatile partitioning between atmosphere and interior. Whereas \citet{werlen_sub-neptunes_2025} reported a maximum interior water content of $\sim 1.5\,\mathrm{wt}\%$, the \revisedtext{fixed-$k_{\rm eq}$ prescription} allows a small fraction of planets to reach somewhat higher total water mass fractions. In our synthetic sample, approximately $3\%$ of planets exceed $1.5\,\mathrm{wt}\%$, and only $\sim 0.3\%$ surpass $5\,\mathrm{wt}\%$, while none attain $10\,\mathrm{wt}\%$.

The enhanced hydrogen storage predicted by our pressure-dependent solubility treatment is conceptually consistent with recent high-pressure experiments \citep{miozzi_experiments_2025, horn_building_2025}, which indicate that significant amounts of hydrogen and water can be incorporated into silicate melts under relevant interior conditions.

\subsection{Silane and Silicon Monoxide}

Silicon-bearing gases provide an additional window into interior–atmosphere exchange. We find that interior–atmosphere equilibration produces large amounts of \ce{SiH4} and \ce{SiO}, with systematic differences between planets formed inside and outside the ice line. The presence of these species is a direct consequence of magma ocean–atmosphere interaction: silicon is supplied from the silicate melt through evaporation reactions, while reduction in hydrogen-rich environments favors the formation of \ce{SiH4}.

Several recent studies have highlighted \ce{SiH4} (and the \ce{SiO}--\ce{SiH4} balance) as a potential tracer of magma ocean--atmosphere interaction under reducing, hydrogen-rich conditions \citep{misener_atmospheres_2023, charnoz_effect_2023, ito_monosilane_2025, hakim_silanemethane_2026}. These works emphasize that the abundance of reduced silicon species is sensitive to redox state, hydrogen partial pressure, and, more generally, the thermochemical structure of the atmosphere. Our population-level results are consistent with this picture and further suggest that the silicon inventory available for outgassing can modulate the \ce{SiH4} yield across formation environments. More detailed forward modeling indicates that the degree to which such deep-atmosphere signatures translate to the observable atmosphere depends on atmospheric structure and additional processes such as condensation and transport \citep{misener_atmospheres_2023, hakim_silanemethane_2026, nixon_magma_2025}.

\subsection{Link to Observations}

Although our framework does not resolve the atmosphere radially, we compare the equilibrium mixing ratios at the atmosphere–magma ocean interface and the resulting bulk atmospheric properties directly to current interpretations of atmospheric observations of various sub-Neptunes. \revisedtext{These equilibrium abundances correspond to the deep atmosphere at the AMOI, where pressures in our population can reach up to $\sim 10\,\mathrm{GPa}$, and should not be interpreted as direct predictions for the observable mbar--bar atmosphere. Transport to observable pressures requires chemical quenching along the pressure--temperature profile, and the resulting quenched abundances can differ substantially from the local AMOI equilibrium values. For example, \citet{werlen_atmospheric_2025} found quench pressures of order $\sim 10\,\mathrm{bar}$ for $K_{zz}=10^7\,\mathrm{cm^2\,s^{-1}}$, much lower than the deep-boundary pressures considered here but still much deeper than the transmission photosphere.} We note that a rigorous interpretation of transmission spectra requires self-consistent pressure--temperature profiles, vertical mixing, cloud formation, and photochemistry. For example, \citet{werlen_atmospheric_2025} and \citet{lee_mineral_2025} showed that weak vertical mixing and silicate cloud chemistry can substantially alter atmospheric abundances relative to deep equilibrium values, highlighting the need for coupled atmospheric structure and spectral modeling.

\subsubsection{TOI-270\,d}

TOI-270 d hosts the so far best-characterized sub-Neptune atmosphere. Using JWST transmission spectra, \citet{benneke_jwst_2024} report strong detections of \ce{CH4}, \ce{CO2}, tentative detections of \ce{H2O} and \ce{CS2}, infer a metallicity of $\sim 225\times$ Solar corresponding to $Z_{\rm atm} = 58^{+8}_{-12}\,\%$, and derive C/O $= 0.47^{+0.16}_{-0.19}$. \ce{NH3} is not detected, with $X_{\ce{NH3}} \lesssim 10^{-4}$. An independent analysis by \citet{holmberg_possible_2024}, that was recently followed up on in \citet{constantinou_toi270d_2025}, likewise favors an elevated mean molecular weight. They find a C/O $< 1$, and a non-detection of \ce{NH3} with a similar upper limit.

In contrast, \citet{felix_competing_2025} find that extended sulfur chemistry permits a sulfur-rich solution with \ce{CS2} mixing ratios of order $10^{-2}$, sulfur enrichment exceeding $10^3\times$ Solar, and C/O $\sim 3$. While this scenario remains metal-rich, it implies a carbon-rich atmosphere and substantially different elemental partitioning, highlighting the current degeneracy between oxygen-rich and sulfur-enhanced interpretations. While formation of this amount of \ce{CS2} cannot be achieved through chemical equilibrium, non-equilibrium processes starting with the abundant carbon-species would lead to its production at high abundance \citep{Moses_sulfur_2024, veillet_sulfur_2025}.

Within our equilibrium population, the inferred metal mass fraction of TOI-270\,d lies within the high-metal-rich tail of planets formed inside the ice line. Our models naturally yield oxygen-rich atmospheres at high metal mass fractions and predict \ce{NH3} abundances of $10^{-9}$--$10^{-8}$, consistent with observational upper limits. Formation outside the ice line cannot be ruled out, which is mainly supported by a possible super-unity C/O (in sulfur-rich retrievals). However, the extreme sulfur enrichment in the sulfur-rich retrievals is not reproduced under equilibrium interior--atmosphere exchange alone.

\subsubsection{K2-18\,b}

K2-18\,b has emerged as a key target for atmospheric characterization of temperate sub-Neptunes. JWST transmission spectra analyzed by \citet{madhusudhan_carbon-bearing_2023} reveal statistically significant detections of \ce{CH4} (5$\sigma$) and \ce{CO2} (3$\sigma$), while \ce{NH3} and \ce{CO} are not detected. The retrieved composition features substantial carbon-bearing species, with \ce{CH4} and \ce{CO2} mixing ratios of order $\sim 10^{-2}$.

A recent reanalysis by \citet{schmidt_comprehensive_2025} confirms the detections of \ce{CH4}, but does not reproduce the \ce{CO2} detection. They did not find significant evidence for \ce{NH3}. While details of the retrieved abundances vary, the overall picture of a carbon-bearing atmosphere with suppressed ammonia remains robust across independent analyses.

With 4 additional observations \citet{hu_k2-18b_2025} managed to definitively detect \ce{CO2} at $X_{\ce{CO2}} \sim 10^{-3}$. Combining their \ce{CO2}/\ce{CH4} ratio of order unity to ten with model predictions from \citet{Yang_mapping_2024}, they infer bulk water contents of $10-25\%$ for the interior of K2-18 b underneath a $100\times$ Solar atmosphere, or, alternatively, the presence of a water-ocean underneath a small atmosphere. Notably, this inference of water is not directly based on their non-detection of \ce{NH3}.

In our equilibrium population, atmospheres with \ce{CH4} and \ce{CO2} mixing ratios of order $10^{-2}$ are naturally produced at high atmospheric metal mass fractions. The predicted \ce{NH3} abundances remain well below current observational upper limits, consistent with its non-detection. Depending on formation location and interior–atmosphere exchange, our models produce both sub- and supersolar C/O ratios, allowing for a broad range of carbon-rich and oxygen-rich solutions consistent with present observational uncertainties. However, our models, as discussed in Section \ref{sec:water_mass_fraction} and in previous work by \citet{werlen_sub-neptunes_2025}, do not reproduce the large water mass fraction suggested by \citet{Yang_mapping_2024} and \citet{hu_k2-18b_2025}.

\subsubsection{GJ 3470 b}
While slightly larger than TOI-270\,d and K2-18\,b,
the Neptune-sized planet GJ 3470 b is one of the few other sub-giant planets with a constrained C/O of $0.35\pm0.10$ as well as a measured metallicity of $\sim 100\times$ solar \citep{beatty_sulfur_2024}. These inferred values are driven by the detections of \ce{H2O}, \ce{CH4}, \ce{CO2}, and \ce{SO2}. The same constraints are used to derive C/S and O/S ratios that are interpreted to suggest formation $10-30$ AU from the host star \citep{Turrini_CNOStrace_2021, crossfield_sulfur_2023}. At the same time, the retrieved nitrogen-bearing species \ce{NH3} and \ce{HCN} remain undetected.

The retrieved abundances from \citet{beatty_sulfur_2024} align fairly well with our equilibrated abundances for both of our formation cases, although formation inside the ice line better describes the \ce{H2O} and \ce{CH4} abundances. Neither scenario describes the \ce{SO2} mixing ratio, although this one is likely to be highly influenced by disequilibrium chemistry as already inferred in \citet{beatty_sulfur_2024}. Our results suggest that the formation location of GJ 3470 b remains uncertain for now, and similarly as described in Section \ref{sec:formation_location} the elemental abundances after equilibration might not represent the abundances after accretion. Additionally, the non-detections of \ce{NH3} and \ce{HCN} are a natural outcome of our equilibration calculations.

\subsubsection{Water World vs.\ Magma Ocean Scenario}

The depletion of atmospheric \ce{NH3} has been proposed as a potential tracer of global water oceans on H$_2$-dominated sub-Neptunes, as ammonia can efficiently dissolve in liquid water and thereby be removed from the observable atmosphere. The detection of \ce{CO2} (and other carbon-bearing species) together with the non-detection of \ce{NH3} has been suggested to be consistent with liquid water–atmosphere equilibrium \citep[e.g.,][]{hu_unveiling_2021,tsai_inferring_2021,madhusudhan_chemical_2023}.

\citet{shorttle_distinguishing_2024} argued that strong nitrogen solubility in reducing silicate melts can likewise suppress atmospheric nitrogen in H$_2$-rich planets undergoing magma ocean equilibration. Under such conditions, nitrogen partitions efficiently into the interior, preventing large atmospheric \ce{NH3} abundances even when the bulk nitrogen inventory is significant. Our equilibrium population models recover the same qualitative behavior: across a broad range of metallicities, redox states, and volatile inventories, nitrogen preferentially partitions into the interior and \ce{NH3} remains subdominant in the atmosphere.

Consistent with this picture, \citet{nixon_magma_2025} demonstrated that magma ocean–atmosphere interactions can reproduce the transmission spectrum of TOI-270\,d within a self-consistent framework, and that \ce{NH3} does not become a dominant atmospheric carrier under magma ocean conditions. The non-detection of \ce{NH3} in both K2-18\,b and TOI-270\,d is therefore fully compatible with interior–atmosphere equilibration in a magma ocean scenario and does not uniquely require a water world interior.

\subsubsection{Other planets}

In contrast to TOI-270 d, K2-18 b, and GJ 3470 b, constraints on most other sub-Neptunes remain limited. Only TOI-421 b has been shown to host a definitively low mean molecular weight atmosphere \citep{Davenport_toi421b_2025}, while most other targets are consistent with at least moderately elevated mean molecular weights. Single-molecule detections in LP 791-18 c \citep{Roy_LP791-18c_2025} and TOI-732 c \citep{Rigby_TOI-732c_2025} provide constraints on metallicity, but do not yet robustly constrain elemental ratios.

A major limitation is the degeneracy between high-altitude clouds and high mean molecular weight atmospheres. In a few cases, this degeneracy has been lifted, revealing high mean molecular weight atmospheres (e.g., GJ 9827 d; \citealt{piaulet_GJ9827d_2024}) or disfavoring cloud-dominated scenarios (e.g., LHS 1140 b; \citealt{Damiano_LHS1140_2024, Cadieux_LHS1140_2024}). Other systems indicate that high metallicity and clouds can coexist, as in GJ 1214 b, where JWST observations confirm an extremely metal-rich atmosphere \citep{kempton_reflective_2023, Schlawin_gj1214_2024, Ohno_gj1214_2025}. However, many observations remain inconclusive, such as emission spectroscopy of GJ 436 b \citep{Mukherjee_GJ436b_2025} and the COMPASS survey targets \citep{Batalha_compass_2023, gordon_compass_2026}.

Our results suggest that elevated to high mean molecular weight atmospheres are a natural and common outcome of magma-ocean equilibration (Figure~\ref{fig:kde_mean_molecular_weight}). While planets formed inside the ice line retain a fraction of low mean molecular weight atmospheres, planets formed outside the ice line are almost exclusively shifted to high mean molecular weight. We emphasize that our results describe the deep atmosphere at the AMOI, whereas transmission spectra probe mbar pressures. Nevertheless, if vertical transport is efficient, the prevalence of elevated mean molecular weights should extend to observable levels, consistent with current observations.

\subsection{Caveats and Future Work}

\subsubsection{Population and disk assumptions.}
The synthetic planet population considered here is restricted to Sun-like host stars and a single underlying disk chemistry. Variations in stellar abundances and disk C/O ratios are therefore not captured. In addition, refractory carbon (e.g., PAHs, soot, or amorphous carbon) in solids is not included. This is consistent with a scenario where interstellar refractory carbon was destroyed during processing in the protoplanetary disk, leading to a chemical \textit{reset} of the solid inventory \citep{anderson_destruction_2017,bergin_tracing_2015}. However, cold \textit{inheritance} of interstellar material would instead require refractory carbon to survive disk processing and be incorporated into planet-forming solids, as observed in primitive outer Solar System material such as comets and carbonaceous chondrites \citep{alexander_nature_2017}. Our predicted atmospheric carbon abundances for this class should thus be regarded as conservative lower limits.

\subsubsection{Thermodynamic uncertainties.}
\revisedtext{Apart from the hydrogen solubility treatment discussed in Section~\ref{sec:h2_solubility}, additional thermodynamic uncertainties remain.} Thermodynamic data for silicate and metal species, particularly for silicate–metal partitioning at high pressure and temperature, remain sparse. We assume ideal mixing in the gas and silicate phases and include activity coefficients only for selected metal species (Appendix~\ref{ap:therm_data}). While internally consistent, this treatment neglects potential non-ideal interactions in other phases. As shown by \citet{werlen_effects_2026}, applying non-ideality corrections to only a subset of species can introduce artificial trends. Progress therefore requires improved high-pressure thermodynamic measurements and comprehensive activity models spanning gas, silicate, and metal reservoirs.

\subsubsection{Photochemical Stability of Silicon-Bearing Species.}
Connecting magma-ocean equilibrium models to observations requires coupling deep-interior chemistry to atmospheric structure and kinetics codes in order to follow abundances to lower pressures and temperatures. Such frameworks are available and include radiative–convective and kinetics models such as HELIOS \citep{malik_helios_2017, malik_self-luminous_2019} and VULCAN \citep{tsai_vulcan_2017}, which allow self-consistent treatment of vertical transport and photochemistry.

Linking the deep atmosphere to the observable atmosphere introduces additional challenges. The photochemistry of silicon-bearing species remains poorly constrained. In particular, reaction pathways and rate coefficients for \ce{SiH4} under relevant temperature and pressure conditions are uncertain \citep[e.g.,][]{ito_monosilane_2025}, limiting robust predictions of its abundance at observable altitudes. Laboratory measurements and dedicated theoretical calculations of silicon reaction networks are therefore essential to assess whether \ce{SiH4} can serve as a reliable tracer of magma–ocean processing.

\subsubsection{Evolution and Atmospheric Escape}

We extract the population from the NGPPS shortly after disk dispersal ($\sim$~4~Myr). The results presented here therefore represent the atmospheric composition at this early evolutionary stage. Subsequent planetary evolution, including atmospheric escape and thermal cooling, is not included in the present framework.

Atmospheric escape is expected to modify the atmospheric composition over time. \revisedtext{The degree to which escape fractionates atmospheric species depends on the escape regime; in hydrodynamic H/He outflows, heavier species can be dragged along with the escaping gas \citep[e.g.,][]{koskinen_escape_2013,koskinen_thermal_2014}. When fractionation does occur,} mass loss can lead to atmospheric enrichment in heavier elements and alter the volatile inventory of the atmosphere. However, significant changes in atmospheric composition are only expected for planets close to the radius valley \citep{cherubim_strong_2024,valatsou_oxygen_2026}.

In addition, the thermal evolution of the planet changes the temperature at the atmosphere--magma ocean interface (AMOI), which in turn affects volatile partitioning between the atmosphere and the interior. Coupled thermal--chemical evolution models show that cooling can lead to the exsolution of volatiles from the interior and modify the atmospheric composition over time \citep{steinmeyer_coupled_2026}. 

\section{Conclusion}
\label{sec:conclusion}

Using a synthetic population of young super-Earths and sub-Neptunes drawn from the New Generation Planetary Population Synthesis shortly after disk dispersal ($\sim$~4~Myr), and post-processed with an extended global chemical equilibrium framework, we have demonstrated that magma ocean--atmosphere equilibration fundamentally reshapes primordial atmospheres. By including sulfur- and nitrogen-bearing species, we show that substantial fractions of N, S, H, and C are redistributed into silicate and metal reservoirs during equilibration. As a consequence, equilibrated atmospheres differ markedly from their accreted states. Nitrogen is strongly depleted from the atmosphere due to dissolution in the silicate melt. Sulfur exhibits compensating trends linked to opposing bulk hydrogen, oxygen, and sulfur abundances across formation environments. Silicon-bearing gases such as \ce{SiH4} and \ce{SiO} emerge as direct tracers of magma ocean exchange. Interior--atmosphere processing systematically modifies elemental ratios, including C/O, and reduces total atmospheric mass fractions through volatile partitioning into the interior.

Formation location relative to the water ice line leaves clear but indirect signatures on atmospheric composition. Planets formed outside the ice line retain enhanced bulk carbon and nitrogen inventories and occupy comparatively narrow C/O and \ce{SiH4} distributions, whereas planets formed inside the ice line exhibit broader compositional spreads and, in many cases, enhanced atmospheric metallicities after equilibration. At the population level, our results demonstrate that observable atmospheric abundances do not directly reflect accreted disk compositions, but instead encode the coupled evolution of bulk composition, redox state, refractory inventory, and interior volatile storage.

A key observational implication concerns nitrogen chemistry. We find that ammonia depletion is a robust and generic outcome of magma ocean equilibration under reducing conditions, as nitrogen preferentially partitions into the interior across wide regions of parameter space. The absence of detectable \ce{NH3} in sub-Neptune atmospheres therefore does not uniquely diagnose the presence of surface liquid water, but is fully consistent with interior--atmosphere exchange in a molten silicate scenario. Interpreting atmospheric measurements without accounting for such chemical processing can bias inferences about planetary structure and formation environment. Coupling deep-interior equilibrium models to atmospheric structure and photochemical frameworks will be essential for translating these equilibrium predictions into robust observational diagnostics and for reliably interpreting transmission and emission spectra obtained with facilities such as \textit{JWST} and the upcoming \textit{ELT} and \textit{Ariel} mission.

%% Please use the acknowledgment and contribution environments. This will 
%% be anonomyized when the "anonymous" style option is used. 
\begin{acknowledgments}
R.B. acknowledges the financial support from the Observatoire de la C\^ote d'Azur via the Poincar\'e fellowship. C.D. acknowledges support from the Swiss National Science Foundation under grant TMSGI2\_211313. This work has been carried out within the framework of the NCCR PlanetS supported by the Swiss National Science Foundation under grant 51NF40\_205606. \revisedtext{We thank the anonymous reviewer for their insightful comments, which greatly helped to improve this study.} We acknowledge the use of large language models (LLMs), including ChatGPT, to improve the grammar, clarity, and readability of the manuscript.
\end{acknowledgments}

\begin{contribution}
A.W. designed the study, wrote the manuscript, and extended the Global Chemical Equilibrium framework to include nitrogen. R.B. provided the synthetic planet population data and contributed to the interpretation of the results and to the manuscript. C.D. supervised the project and contributed to the manuscript. L.F. contributed to the connection with exoplanet observations. A.S. and A.W. implemented sulfur chemistry into the original equilibrium framework, which served as the basis for this study. All authors read and provided comments for the manuscript.
\end{contribution}




\section*{ORCID iDs}

\noindent 
Aaron Werlen \orcidlink{0009-0005-1133-7586} \href{https://orcid.org/0009-0005-1133-7586}{0009-0005-1133-7586} \\
Remo Burn \orcidlink{0000-0002-9020-7309} \href{https://orcid.org/0000-0002-9020-7309}{0000-0002-9020-7309} \\
Caroline Dorn \orcidlink{0000-0001-6110-4610} \href{https://orcid.org/0000-0001-6110-4610}{0000-0001-6110-4610} \\
Lukas Felix \orcidlink{0009-0004-1292-3969} \href{https://orcid.org/0009-0004-1292-3969}{0009-0004-1292-3969} \\
Annika Salmi \orcidlink{0009-0007-2917-2390} \href{https://orcid.org/0009-0007-2917-2390}{0009-0007-2917-2390}

\appendix
\twocolumngrid

\section{Chemical Network}\label{ap:chem_network}

The chemical network employed in this study follows the general framework introduced by \cite{schlichting_chemical_2022}. Consistent with \cite{werlen_atmospheric_2025, werlen_sub-neptunes_2025}, carbon is treated as an explicit component of the metal phase. Building on this existing framework, we extend the network to include sulfur, nitrogen, and iron redox chemistry. The resulting system is defined by a set of independent basis reactions spanning silicate, metal, and gas phases. Chemical exchange is permitted both within individual reservoirs and across phase boundaries.

Below we list the specific reaction network adopted in this study. Chemical species are distinguished by their host phase, with gas-phase species denoted by the subscript $g$, silicate-phase species by $s$, and metal-phase species by $m$.

Reactions internal to the silicate phase are given by:

\begin{equation}
    \ce{Na2SiO3_{,\text{s}} \rightleftharpoons Na2O_{\text{s}} + SiO2_{,\text{s}}} \tag{R1}
\end{equation}

\begin{equation}
    \ce{MgSiO3_{,\text{s}} \rightleftharpoons MgO_{\text{s}} + SiO2_{,\text{s}}} \tag{R2}
\end{equation}

\begin{equation}
    \ce{FeSiO3_{,\text{s}} \rightleftharpoons FeO_{s} + SiO2_{,\text{s}}} \tag{R3}
\end{equation}

\begin{equation}
    \ce{2FeO_{\text{s}} + \tfrac{1}{2}O2_{,\text{g}}} \rightleftharpoons \ce{2FeO_{1.5,\text{s}}} \tag{R4}
\end{equation}

Reactions governing exchange between the metal and silicate phases include:

\begin{equation}
    \ce{FeO_{\text{s}} + \tfrac{1}{2}Si_{\text{m}} \rightleftharpoons Fe_{\text{m}} + \tfrac{1}{2}SiO2_{,\text{s}}} \tag{R5}
\end{equation}

\begin{equation}
    \ce{O_{\text{m}} + \tfrac{1}{2}Si_{\text{m}} \rightleftharpoons \tfrac{1}{2}SiO2_{,\text{s}}} \tag{R6}
\end{equation}

\begin{equation}
    \ce{2H_{\text{m}} \rightleftharpoons H2_{,\text{s}}} \tag{R7}
\end{equation}

\begin{equation}
    \ce{Si_{\text{m}} + 2H2O_{\text{s}} \rightleftharpoons SiO2_{,\text{s}} + 2H2_{,\text{s}}} \tag{R8}
\end{equation}

\begin{equation}
    \ce{CO_{\text{s}}\rightleftharpoons C_{\text{m}} + O_{\text{m}}}  \tag{R9}
\end{equation}

\begin{equation}
    \ce{FeS_{\text{s}} \rightleftharpoons Fe_{\text{m}} + S_{\text{m}}} \tag{R10}
\end{equation}

Equilibria among gas-phase species are described by:

\begin{equation}
    \ce{CO_{\text{g}} + \tfrac{1}{2}O2_{,\text{g}} \rightleftharpoons CO2_{,\text{g}}} \tag{R11}
\end{equation}

\begin{equation}
    \ce{CH4_{,\text{g}} + \tfrac{1}{2}O2_{,\text{g}}} \rightleftharpoons \ce{2H2_{,\text{g}} + CO_{\text{g}}} \tag{R12}
\end{equation}

\begin{equation}
    \ce{H2_{,\text{g}} + \tfrac{1}{2}O2_{,\text{g}} \rightleftharpoons H2O_{\text{g}}} \tag{R13}
\end{equation}

\begin{equation}
    \ce{SO2_{,\text{g}} + H2_{,\text{g}} \rightleftharpoons H2S_{\text{g}} + O2_{,\text{g}}} \tag{R14}
\end{equation}

\begin{equation}
    \ce{2NH3_{,\text{g}}} \rightleftharpoons \ce{3H2_{,\text{g}} + N2_{,\text{g}}} \tag{R15}
\end{equation}

\begin{equation}
    \ce{NH3_{,\text{g}} + CH4_{,\text{g}}} \rightleftharpoons \ce{HCN_{\text{g}} + 3H2_{,\text{g}}} \tag{R16}
\end{equation}

Exchange between the magma ocean and the atmosphere is represented by:

\begin{equation}
    \ce{FeO_{\text{s}} \rightleftharpoons Fe_{\text{g}} + \tfrac{1}{2}O2_{,\text{g}}} \tag{R17}
\end{equation}

\begin{equation}
    \ce{MgO_{\text{s}} \rightleftharpoons Mg_{\text{g}} + \tfrac{1}{2}O2_{,\text{g}}} \tag{R18}
\end{equation}

\begin{equation}
    \ce{SiO2_{,\text{\text{s}}} \rightleftharpoons SiO_{\text{g}} + \tfrac{1}{2}O2_{,\text{g}}} \tag{R19}
\end{equation}

\begin{equation}
    \ce{Na2O_{\text{s}}} \rightleftharpoons \ce{2Na_{\text{g}} + \tfrac{1}{2}O2_{,\text{g}}} \tag{R20}
\end{equation}

\begin{equation}
    \ce{H2_{,\text{s}} \rightleftharpoons H2_{,\text{g}}} \tag{R21}
\end{equation}

\begin{equation}
    \ce{N2_{,\text{s}} \rightleftharpoons N2_{,\text{g}}} \tag{R22}
\end{equation}

\begin{equation}
    \ce{H2O_{\text{s}} \rightleftharpoons H2O_{\text{g}}} \tag{R23}
\end{equation}

\begin{equation}
    \ce{CO_{\text{s}} \rightleftharpoons CO_{\text{g}}} \tag{R24}
\end{equation}

\begin{equation}
    \ce{CO2_{,\text{s}} \rightleftharpoons CO2_{,\text{g}}} \tag{R25}
\end{equation}

\begin{equation}
    \ce{2FeSO4_{,\text{s}}} \rightleftharpoons \ce{2FeO_{\text{s}} + 2SO2_{,\text{g}} + O2_{,\text{g}}} \tag{R26}
\end{equation}

\begin{equation}
    \ce{3H2O_{\text{s}} + FeS_{\text{s}}} \rightleftharpoons \ce{3H2_{,\text{s}} + FeO_{\text{s}} + SO2_{,\text{g}}} \tag{R27}
\end{equation}

Any reaction that can be expressed as a linear combination of these basis reactions is implicitly allowed, such that the accessible chemical space is not restricted to the explicit reactions listed above.

Chemical equilibrium is obtained by solving the mass-action conditions for all basis reactions, following \citet{schlichting_chemical_2022}. For each reaction, equilibrium is enforced via

\begin{equation}\label{eq:chemical_equilibrium}
    \sum_i \nu_i \ln x_i + \left[\frac{\Delta \hat{G}^\circ_{\mathrm{rxn}}}{RT} + \sum_g \nu_g \ln(P/P^\circ)\right] = 0,
\end{equation}

\noindent where $x_i$ is the mole fraction of species $i$ in its respective phase and $\nu_i$ are stoichiometric coefficients. $\Delta \hat{G}^\circ_{\mathrm{rxn}}$ denotes the standard Gibbs free energy change, $R$ is the ideal gas constant, and $T$ is temperature. The summation over $g$ includes gas-phase species only and introduces the explicit dependence on the atmosphere--magma ocean interface (AMOI) pressure $P$, with reference pressure $P^\circ = 1$~bar.

The system is closed by imposing normalization constraints such that mole fractions in each phase sum to unity, and by enforcing elemental conservation across the gas, silicate, and metal reservoirs. Following \citet{schlichting_chemical_2022}, both the mole fractions and the total number of moles in each phase are treated as free variables, and the AMOI pressure is determined self-consistently within the coupled solution.

Global equilibrium is solved using the numerical method described by \citet{schlichting_chemical_2022}, with additional algorithmic improvements detailed in \citet{grimm_new_2026}.

\section{Thermodynamic Data}\label{ap:therm_data}

\subsection{Gibbs free energy}

Thermodynamic data for the majority of species considered in this work are described in detail in the appendix of \citet{schlichting_chemical_2022}. This includes all species except nitrogen- and sulfur-bearing species, ferric iron ($\ce{FeO}_{1.5}$), hydrogen dissolved in the silicate melt, and carbon in the metal phase, which are treated separately here.

Thermodynamic data for sulfur- and nitrogen-bearing gas-phase species are taken from the NIST Chemistry WebBook\footnote{\url{https://webbook.nist.gov/chemistry/}} \citep{NIST}. Gibbs free energy data for $\ce{FeO}_{1.5}$ in the silicate melt are obtained from the MELTS thermodynamic model \citep{ghiorso_chemical_1995, asimow_algorithmic_1998}, accessed through \texttt{alphaMELTS}~2.0 \citep{smith_adiabat_1ph_2005} using the Python wrapper\footnote{\url{https://github.com/magmasource/alphaMELTS}} at 1 bar, and are fitted using Chebyshev polynomials. The resulting fit is

\begin{equation}
\begin{split}
G_{\mathrm{FeO}_{1.5}}(T,1~\mathrm{bar}) &=
\sum_{i=0}^{5} a_i\,T_i(x_T), \\
x_T &= \frac{T-2523.15}{750}.
\end{split}
\end{equation}

\noindent where $T_i(x)$ are Chebyshev polynomials of the first kind, defined by

\begin{equation}
\begin{aligned}
&T_0(x) = 1, \hspace{1em} T_1(x)=x,\\
&T_n(x) = 2x\,T_{n-1}(x)-T_{n-2}(x).
\end{aligned}
\end{equation}

The fitted coefficients are

\begin{equation}
\begin{aligned}
a_0&=-7.6853\times10^{5},\hspace{1em} & a_3&=\phantom{-}2.7177\times10^{2},\\
a_1&=-1.7706\times10^{5},\hspace{1em} & a_4&=-4.8078\times10^{1},\\
a_2&=-6.9694\times10^{3},\hspace{1em} & a_5&=-7.2718\times10^{-1}.
\end{aligned}
\end{equation}

The Gibbs free energy of \ce{H2} dissolved in the silicate melt is derived from a combined fit to experimental measurements by \citet{hirschmann_solubility_2012} and \citet{gilmore_coreenvelope_2026}, following the procedure described in \citet{werlen_effects_2026}. For carbon in the metal phase, metal--silicate partitioning coefficients from \citet{blanchard_metalsilicate_2022} are adopted, as implemented in \citet{werlen_atmospheric_2025}. Thermodynamic data for sulfur in the metal phase are adopted from \citet{calvo_accretion_2026}. The pressure dependence of the sulfur partitioning coefficients is neglected, as we find that it has only minor effects on the results, and this choice maintains consistency with the treatment of other metal species. We also neglect compositional dependencies associated with species not included in the present model.

Thermodynamic data for \ce{N2} dissolved in the silicate melt are obtained from the enthalpy, entropy, and volume parameterization of the nitrogen solubility reaction reported by \citet{bernadou_nitrogen_2021} (see their Table~6).

\subsection{Non-ideality treatment}

Non-ideal mixing in the metal phase is treated by replacing mole fractions with activities, $a_i = \gamma_i x_i$, for selected species. Activity coefficients for Si and O in the metal are implemented following \citet{badro_core_2015} and \citet{young_earth_2023}. Non-ideal mixing for C in the metal is included using the formulation of \citet{fischer_carbon_2020}.

We assume both the gas and silicate melt phases to behave ideally. While activity coefficients are known to be important for specific volatile–melt interactions, comprehensive and internally consistent activity models for the full set of silicate species considered here are currently sparse. Moreover, applying non-ideality corrections to only a subset of phases has been shown to introduce artificial shifts in equilibrium partitioning and lead to inaccurate or misleading trends \citep{werlen_effects_2026}. For this reason, we adopt an internally consistent ideal treatment for the gas and silicate phases in the present work.

%% For this sample we use BibTeX plus aasjournalv7.bst to generate the

\bibliography{references}{}
\bibliographystyle{aasjournalv7}



\end{document}